\documentclass[twocolumn,amsmath,amssymb]{revtex4}
\usepackage{graphicx}
\usepackage{dcolumn}
\usepackage{bm}
\begin{document}

\title{Emergent Geometry Fluctuation in Quantum Confined Electron Systems} 
\author{Areg Ghazaryan and Tapash Chakraborty}
\affiliation{Department of Physics and Astronomy,
University of Manitoba, Winnipeg, Canada R3T 2N2}

\date{\today}
\begin{abstract}
The intrinsic geometric degree of freedom that was proposed to determine the 
optimal correlation energy of the fractional quantum Hall states, is analyzed  
for quantum confined planar electron systems. One major advantage in this case
is that the role of various unimodular metrics resulting from the absence of rotational
symmetry can be investigated independently or concurrently. For interacting electrons 
in our system, the confinement metric due to the anisotropy shifts the minimum of the 
ground state and the low-lying excited states from the isotropic case much more strongly 
than the corresponding shift due to the unimodular Galilean metric. Implications of these 
results for possible observation of higher Landau level filling fractions have been
elucidated.
\end{abstract}
\maketitle

In a two-dimensional electron gas (2DEG) subjected to a strong perpendicular magnetic field,
the ground state corresponding to the $\frac13$ filling factor of the lowest Landau
level \cite{fqhe,FQHE_book} is described by the celebrated Laughlin wave function \cite{laughlin}.
While investigating the origin of the success of that wave function,
Haldane \cite{haldane_11} recently realized that, contrary to the popular belief,
there is a hidden geometrical fluctuation corresponding to the anisotropic correlation
hole around each electron in the system. This anisotropy occurs either in the presence of 
an anisotropic interaction or the band mass anisotropy that gives rise to a unimodular 
(area preserving) Galilean metric \cite{note_1}. In fact, the interaction metric is not necessarily
congruent to the Galilean metric. Interestingly, the latter metric was shown to be 
equivalent \cite{haldane_2_12} to what one obtains in the case of a magnetic field that is
tilted \cite{FQHE_book,tilted} from the direction perpendicular to the electron plane. Consequent
to this theory of Haldane, there were a few numerical studies reported in the literature
\cite{haldane_12,fuchun,yang,vadim}, that explored the influence of the Coulomb or
Galilean metric corresponding to a particular anisotropic Hamiltonian. In these approaches
the correlation metric was taken to interpolate \cite{haldane_11} between the two other 
metrics. It provides the variational parameter one needs to minimize the correlation
energy in the Laughlin state. The lowest Landau level fractional quantum Hall (FQH) states
were found to be robust against variation of the anisotropy introduced through the intrinsic metric,
while the FQH states at higher Landau levels are susceptible to compressible - incompressible
phase transitions due to the anisotropy \cite{haldane_2_12}. In the present work, we report on our 
study of the quantum-confined electron systems \cite{confined} where we introduce various unimodular
metrics that can be varied independently or conjunctionally, thereby providing information about 
their influence on the energy spectra. Our results indicate that, in a single electron system the 
mass anisotropy and the confinement anisotropy will generate identical effects. In the case of 
interacting electrons in the system, the confinement anisotropy shifts the minimum of the ground 
state and the low-lying excited states from the isotropic case much more strongly than the 
corresponding shift due to the mass anisotropy alone. A suitable combination of these two unimodular
metrics would perhaps generate a more pronounced FQHE in experiments at higher Landau level filling 
factors.

We consider the two-dimensional electron gas subjected to a perpendicular magnetic field 
and in a parabolic confinement. The many-body Hamiltonian can then be written in the form
\begin{equation}\label{MBHamiltonian}
\mathcal{H}=\sum_i^N\mathcal{H}_i^e+\tfrac12 \sum_{i,j}^N V^{}_{ij},
\end{equation}
where $\mathcal{H}_i^e$ is a one electron Hamiltonian which with the inclusion 
of the effective mass and the confinement anisotropy is written as
\begin{align}\label{singleH}
\mathcal{H}^e &={\frac1{2m^{}_e}}\left[\left(\Pi^{}_x\right)^2/\alpha^{}_\mu+
\alpha^{}_\mu\left(\Pi^{}_y\right)^2\right]\nonumber \\
&\qquad +{\tfrac12}m^{}_e\omega_0^2\left(\alpha^{}_{\rm C}x^2+y^2/\alpha^{}_{\rm C}\right)+
\tfrac12g\mu^{}_{\rm B} B\sigma^{}_z.
\end{align}
Here $\bf \Pi=\bf p-\frac ec \bf A$, $\omega^{}_0$ is the confinement potential strength for
the isotropic case, while $\alpha^{}_\mu$ and $\alpha^{}_{\rm C}$ are the mass and 
confinement anisotropy parameters respectively. We employ the symmetric gauge vector potential 
${\bf A}= \frac12(-y,x,0)B$. The third term on the right hand side of Eq.~(\ref{singleH}) 
is the Zeeman energy. The second term in Eq.~(\ref{MBHamiltonian}) is the 
Coulomb interaction which in the case of the anisotropic dielectric tensor has the form
\begin{equation}
V^{}_{ij}=\frac{e^2}{\varepsilon \sqrt{\alpha^{}_{\rm I}(x^{}_i-x^{}_j)^2+(y^{}_i-y^{}_j)^2/
\alpha^{}_{\rm I}}},
\end{equation}   
where $\alpha^{}_{\rm I}$ is the interaction anisotropy parameter, and the directions of 
$\hat{x}$ and $\hat{y}$ are along the principal axes of the dielectric tensor. 
We first decouple the one-particle Hamiltonian into three parts
\begin{align}
&\mathcal{H}^{}_\Lambda=\frac1{2m^{}_e\alpha^{}_\mu}\left(p_x^2+\Omega^2_xx^2\right)+
\frac{\alpha^{}_\mu}{2m^{}_e}\left(p_y^2+\Omega^2_yy^2\right), \nonumber \\
&\mathcal{H}^{}_{\rm Z}=\tfrac12g\mu^{}_{\rm B} B\sigma^{}_z, \nonumber \\
&\mathcal{H}^{}_{\rm R}=\tfrac12\omega^{}_{\rm c}\left(\alpha^{}_\mu xp^{}_y-yp^{}_x/\alpha^{}_\mu\right),
\nonumber
\end{align}
where $\mathcal{H}^{}_\Lambda$ describes the two-dimensional spinless harmonic 
oscillator with mass anisotropy in $\hat{x}$ and $\hat{y}$ directions. We have 
introduced the cyclotron frequency $\omega^{}_{\rm c}=|e|B/m^{}_e c$ and the oscillator frequencies 
$\Omega^2_x=m^{2}_e\alpha_\mu^2(\frac{\alpha^{}_{\rm C}}{\alpha^{}_\mu}\omega_0^2+\frac14
\omega_{\rm c}^2)$, $\Omega^2_y=\frac{m^{2}_e}{\alpha_\mu^2}(\frac{\alpha^{}_\mu}{\alpha^{}_{\rm C}}
\omega_0^2+\frac14\omega_{\rm c}^2)$. The eigenstates $|\lambda\rangle$ of 
$\mathcal{H}^{}_\Lambda$ are the direct products $|n_x^\lambda\rangle|n_y^\lambda\rangle$ 
of two harmonic oscillator states represented by the quantum numbers $n_{x,y}^\lambda$. 
Inclusion of the Zeeman term $\mathcal{H}^{}_{\rm Z}$ is done by multiplying the states 
$|\lambda\rangle$ by the eigenstates of the Pauli spin matrix $\sigma^{}_z$. Finally, 
the $\mathcal{H}^{}_{\rm R}$ part of the Hamiltonian mixes the states with different 
quantum numbers $n_{x,y}^\lambda$ and therefore the spectra of the single-electron 
Hamiltonian should be obtained by employing the diagonalization procedure using the 
eigenstates of $\mathcal{H}^{}_\Lambda$ and $\mathcal{H}^{}_{\rm Z}$ as the basis.          

In order to evaluate the energy spectrum of the confined electron system we need to diagonalize 
the matrix of the Hamiltonian in Eq.~(\ref{MBHamiltonian}) in a basis of the Slater determinants 
constructed from the single-electron eigenstates of the Hamiltonian in Eq.~(\ref{singleH}). 
To calculate the two-body Coulomb interaction matrix elements we use the procedure of 
Fourier transformation outlined previously \cite{Avetisyan_2}. The Fourier transform of the
anisotropic interaction is also anisotropic and has the form
\begin{equation}
\tilde{V}^{}_{\rm C}(\mathbf{k})=\frac{2\pi e^2}{\varepsilon \sqrt{k_x^2/\alpha^{}_{\rm I}
+k_y^2\alpha^{}_{\rm I}}}.
\end{equation} 
Our numerical studies were carried out for the two-dimensional InAs system using the 
following parameters: $m^{}_e=0.042m^{}_0$, $g^{}_e=-14$, $\varepsilon=14.6$, $\omega_0=4$ meV. 
We have considered the variation of all anisotropy parameters $\alpha^{}_\mu, 
\alpha^{}_{\rm C}$ and $\alpha^{}_{\rm I}$ in the range of $0.2-2$. The value of 
$\alpha^{}_\mu= \alpha^{}_{\rm C} = \alpha^{}_{\rm I} = 1$ correspond to the isotropic case.
The results for the more popular GaAs system are qualitatively similar to the present case. 

\begin{figure}
\includegraphics[width=6cm]{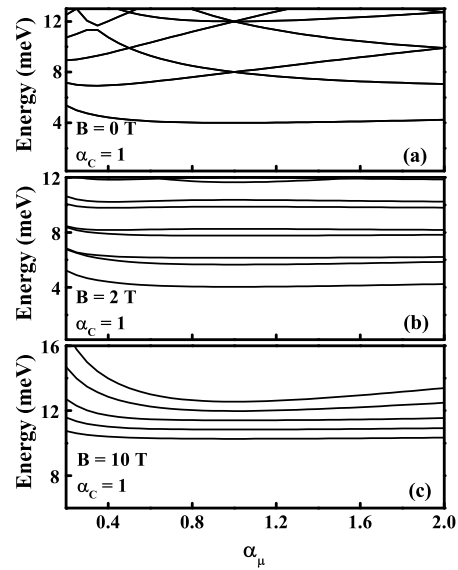}
\caption{\label{fig:OneElectron_Edepam} Dependence of the low-lying energy levels 
of a single electron system on the mass anisotropy parameter $\alpha^{}_\mu$ for various 
values of the magnetic field strength $B$. The confinement is taken to be isotropic.
For $B=0$ all states are Kramer doublets.}
\end{figure}

Let us first discuss about the one electron case. In Fig.~\ref{fig:OneElectron_Edepam} the 
dependence of low-lying energy levels of a one-electron system on the mass anisotropy 
parameter $\alpha^{}_\mu$ is shown for various values of the magnetic field strength $B$. 
The confinement parameter is taken to be $\alpha^{}_{\rm C}=1$ (isotropic confinement 
potential). The spectrum is clearly symmetric under the transformation $\alpha^{}_\mu 
\rightarrow 1/\alpha^{}_\mu$ and the minimum for the ground state appears in the isotropic 
case. Inclusion of the mass anisotropy brakes the rotational symmetry of the system
which results in the lifting of the degeneracies for the excited states in the case of $B=0$. 
It should be noted that the dependence of the energy levels on the confinement parameter 
$\alpha^{}_{\rm C}$ for the case of $\alpha^{}_\mu=1$ is exactly the same as in 
Fig.~\ref{fig:OneElectron_Edepam}. This can be explained directly from the one-electron 
Hamiltonian of Eq. (\ref{singleH}), by making a rescaling of the coordinates $x \rightarrow 
x/\sqrt{\alpha^{}_\mu}, \, y \rightarrow y\sqrt{\alpha^{}_\mu}$. In that case the mass anisotropy 
will be transferred to the confinement anisotropy, with the direction of the hard and easy 
axes interchanged as compared to the anisotropy introduced by $\alpha^{}_{\rm C}$. The 
interchange of the hard and easy axes has no effect on the energy spectrum of the system.
Therefore for the one-electron case it makes no difference whether the symmetry of the 
system is broken by inducing the mass anisotropy or the confinement anisotropy. This might have
important consequences in actual experiments involving anisotropic electron systems. 

\begin{figure}
\includegraphics[width=6cm]{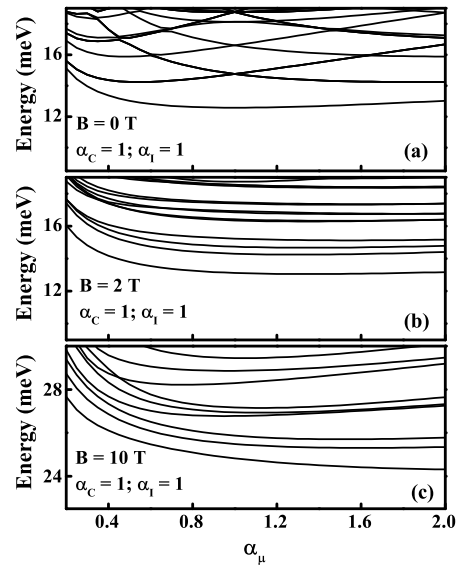}
\caption{\label{fig:TwoElectron_Edepam} Dependence of the low-lying energy levels 
of a two-electron system on the mass anisotropy parameter $\alpha^{}_\mu$ for various 
values of the magnetic field strength $B$. The confinement and the Coulomb interaction 
are taken to be isotropic.}
\end{figure}

In Fig.~\ref{fig:TwoElectron_Edepam} we present the magnetic field dependence of the low-lying 
energy levels of a two-electron system on the mass anisotropy parameter $\alpha^{}_\mu$. 
The confinement and the Coulomb interaction are considered to be isotropic ($\alpha^{}_{\rm C}=1$ 
and $\alpha^{}_{\rm I}=1$). In the absence of an external magnetic field, this anisotropy parameter 
dependence strongly resembles that of the one-electron case. However, for the case of $B=2$T or 
$10$T, the minimum of the ground state and the excited states are clearly shifted to higher values 
of the anisotropy parameter, and the case of $\alpha^{}_\mu=1$ does not correspond to a special 
point. It should be pointed out that, interestingly a similar behavior was reported earlier for the 
FQHE state with filling factor $\nu=1+1/3$ \cite{haldane_2_12} from theoretical studies of 
8 electrons in a periodic rectangular geometry. We have shown previously \cite{madhav,Avetisyan_1} 
that in an ansiotropic system and for the one-electron case the effect of the magnetic field 
is to rotate the directions of the oscillator motion from $\hat{x}$ and $\hat{y}$ directions. 
By making the same rescaling as for the one-electron case mentioned above, the inclusion of mass 
anisotropy (or any one anisotropy) for the many-electron case can be transferred to the
anisotropies of the parabolic confinement and the Coulomb interaction (or to the other two
anisotropies), both having the same hard axis, as long as the metrics are diagonal. The change 
in anisotropy of the confinement potential also affects the angle of rotation of the oscillator 
motion induced by the magnetic field. Hence the shift of the minimum of the ground and excited 
states to higher anisotropy values is an interplay between the rotated oscillator motion which is 
determined by the magnetic field strength, the confinement anisotropy, and the Coulomb interaction 
anisotropy.

\begin{figure}
\includegraphics[width=6cm]{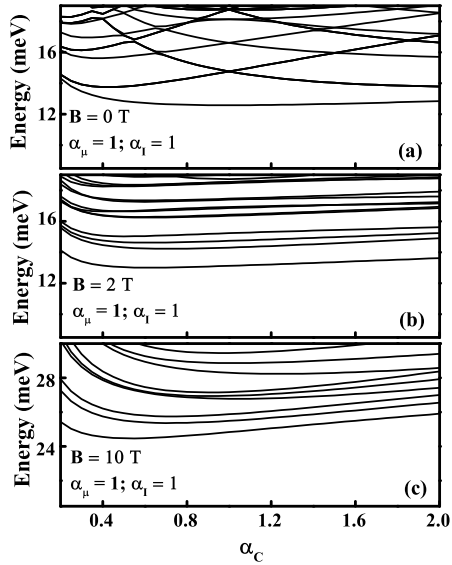}
\caption{\label{fig:TwoElectron_Edepac} Dependence of the low-lying energy levels 
of a two-electron system on the confinement anisotropy parameter $\alpha^{}_{\rm C}$ for various 
values of the magnetic field strength $B$. The mass and the Coulomb interaction are 
taken to be isotropic.}
\end{figure}  

In Fig.~\ref{fig:TwoElectron_Edepac}, we present the dependence of the low-lying 
energy levels of a two-electron system on the confinement anisotropy parameter 
$\alpha^{}_{\rm C}$ for various values of the magnetic field strength $B$. The mass and the Coulomb 
interaction are taken to be isotropic ($\alpha^{}_\mu=1$ and $\alpha^{}_{\rm I}=1$). Here we notice
that, again for the case of the non-zero magnetic field the minimum of the ground and the 
excited states are shifted to a lower value of the corresponding anisotropy parameter. It should 
be pointed out that in this case the shifting of the minimum is more pronounced than in the case 
of the mass anisotropy. Hence the sole inclusion of the confinement anisotropy has a more 
profound effect on the energy levels of the system than the inclusion of the mass anisotropy alone.
The implication of this results for experimental observation of the FQHE states will
be elaborated below.

\begin{figure}
\includegraphics[width=6cm]{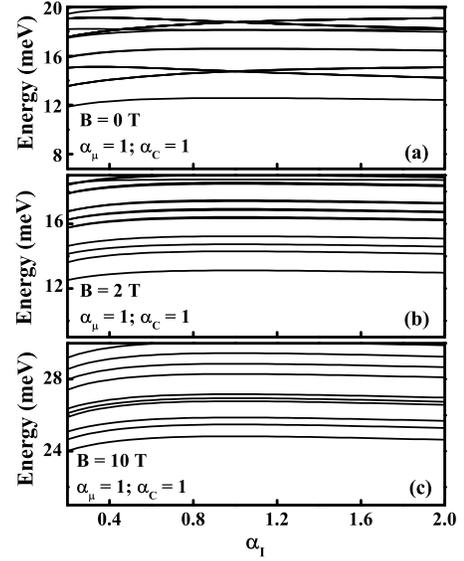}
\caption{\label{fig:TwoElectron_Edepai} Dependence of the low-lying energy levels of a two-electron
system on the interaction anisotropy parameter $\alpha^{}_{\rm I}$ for various values of the 
magnetic field strength $B$. The mass and the confinement are taken to be isotropic.}
\end{figure}

Finally, we consider the case of the interaction anisotropy. The dependence of the low-lying 
energy levels of a two-electron system on the interaction anisotropy parameter $\alpha^{}_{\rm I}$ 
for various values of the magnetic field strength $B$ is presented in 
Fig.~\ref{fig:TwoElectron_Edepai}. The mass and the confinement potential are considered here to 
be isotropic ($\alpha^{}_\mu=1$ and $\alpha^{}_{\rm C}=1$). Here we notice that the energy levels 
have the maximum at the isotropy point for all values of the magnetic field. Due to the isotropy 
of the mass and the confinement potential, the one-electron wave function possesses the rotational 
symmetry. As a consequence, the Coulomb interaction is more pronounced when it is symmetric and 
the electrons interact equivalently in all directions. We can then conclude that the isotropic 
interaction case should have the maximum energy compared to the case of the anisotropic 
interaction.  
   
If the band mass anisotropy is inherent in the semiconductor system considered here and not
induced by an external means, such as the tilted magnetic field, one may argue that the same
mass anisotropy should be applied also to the confinement potential strength parameter, where
the electron effective mass also appears. In fact, systems with internal band mass
anisotropy were suggested by the recent experiments on AlAs two-dimensional electron systems
\cite{Shayegan}. For the one-electron system if both the kinetic term and the parabolic
confinement in the Hamiltonian in Eq.~(\ref{singleH}) possess the same anisotropy, then
using the coordinate rescaling mentioned above it can be shown that these two metrics are
congruent, so that the rotational symmetry is preserved. As for two-electron case using the same
coordinate rescaling in the Hamiltonian (\ref{MBHamiltonian}) this anisotropy can be transfered
to the anisotropy of the Coulomb interaction. If any other anisotropy is imposed on the confinement 
potential by an external means, the anisotropy of the Coulomb interaction will surely move the minimum 
of the dependence of energy levels on the additional confinement anisotropy parameter even farther 
away from the isotropic case, due to the maximum observed for the Coulomb interaction anisotropy 
in the isotropic case.

Experimentally, the tilt-induced anisotropy in the FQHE states have been well studied in recent 
literature \cite{tilt_aniso,Xia} in addition to its earlier success in exploring spin polarizations 
\cite{tilted_spin,jim_old} in the lowest Landau level \cite{tilted}. In a tilted magnetic field 
a stable FQHE state for the filling factor $\nu=\frac73 (= 2+\frac13)$ has been recently reported 
\cite{Xia}. These authors pointed out that albeit the presence of the quantized Hall plateau, the 
longitudinal resistance possesses strong temperature-dependent anisotropy. Theoretically, it was 
shown that the effective mass tensor of the 2DEG can be tuned by the tilted magnetic field 
\cite{haldane_2_12}, and therefore the observation of this special phase at $\nu=7/3$ can be related
to the FQHE with mass anisotropy. It should be pointed out that the tilted magnetic field couples 
the planar motion of the electrons in a 2DEG with the perpendicular motion, rendering the electron 
dynamics in the FQHE states highly non-trivial. Perhaps the mass anisotropy (or as suggested
above, a combination of the mass anisotropy and the confinement anisotropy) will provide a better 
route to study anisotropic states in the FQHE experiments.   

In summary, we have shown that for a two-electron system in a parabolic confinement under the 
influence of a magnetic field, the confinement anisotropy has the similar effect on the system as 
does the mass anisotropy. We have also pointed out that in the case of the confinement anisotropy 
the shifting of the minimum of the ground state and low-lying excited states to lower values of the 
corresponding anisotropy parameter is more pronounced than that for the mass anisotropy. As we have 
mentioned above, a similar kind of shifting of the minimum from the isotropic case was also reported
theoretically for the energy spectrum of 2DEG at filling factor $\nu=1+1/3$. Although these two
calculations are not directly related, the similarity of this shift of the minimum helps us
make a prediction that in the FQHE experiments if one introduces a confinement potential in the 
2DEG, the anisotropy of that confinement potential will have a similar dramatic effect as that of 
the tilted magnetic field. We therefore believe that the simultaneous use of both the tilted 
magnetic field and the confinement anisotropy would strongly enhance the resultant anisotropy than 
the tilt-induced anisotropy alone in the system. That would most likely make the observation of 
the anisotropic FQHE phases even more pronounced. A similar study of larger electron systems will 
be the subject of our future publications. 
   
The work has been supported by the Canada Research Chairs Program of the Government of Canada. 
We thank Pekka Pietil\"ainen and Vadim Apalkov for helpful suggestions.


\begin{thebibliography}{99}
\bibitem{note_1}
The ``Galilean metric" was derived \cite{haldane_11} from the effective mass
tensor $m^{}_{ab}=m\left(\omega^*_a\omega^{}_b+\omega^*_b\omega^{}_a\right)
=m g^{}_{ab}$ in terms of a complex vector $\omega$, and the effective mass 
$m$ that is related to the cyclotron frequency. It is a positive-definite metric 
tensor with det $g=1$ (a unimodular metric).
\end{thebibliography}
\end{document}